\documentclass[a4paper,11pt]{amsart}
\usepackage[latin1]{inputenc}
\usepackage{amsfonts} 
\usepackage{amsmath} 
\usepackage{amssymb}
\usepackage{amsthm}
\usepackage{latexsym} 
\usepackage{mathrsfs}
\usepackage{psfrag}
\usepackage{graphicx}
\usepackage{indentfirst}
\usepackage{setspace,layout}

\usepackage{color}
\usepackage[
  color, 
  final]
 {showkeys}
\definecolor{refkeybis}{gray}{.65}
\definecolor{labelkeybis}{gray}{.65}
{\makeatletter
 \def\SK@refcolor{\color{refkeybis}}
 \def\SK@labelcolor{\color{labelkeybis}}}

\newcommand{\R}{\mathcal{R}}

\newcommand{\N}{\mathbb{N}}

\newcommand{\F}{\mathbb{F}}

\newcommand{\ord}[1]{{\rm ord}(#1)}

\newcommand{\A}{\alpha}

\newtheorem{teo}{Theorem}
\newtheorem{lem}[teo]{Lemma}
\newtheorem{cor}[teo]{Corollary}
\newtheorem{prop}[teo]{Proposition}
\newtheorem{defi}[teo]{Definition}

\theoremstyle{remark}
\newtheorem*{oss}{Remark}

\newenvironment{dem}{\begin{proof}[Proof]} {\end{proof}}

\begin{document}

\title{Computation of the Weight Distribution of CRC Codes.}

\author{Felice Manganiello}
\address{Mathematics Institute\\ Winterthurerstr. 190\\ CH - 8057 Z\"urich}
\email{felice.manganiello@math.unizh.ch}

\maketitle

\begin{abstract}
In this article, we illustrate an algorithm for the computation of the weight
distribution of CRC codes. The recursive structure of
CRC codes will give us an iterative way to compute the weight
distribution of their dual codes starting from just some
``representative'' words. Thanks to MacWilliams Theorem, the
computation of the weight distribution of dual codes can be easily
brought back to that of CRC codes. 
This algorithm is a good alternative to the standard algorithm that
involves listing every word of the code. 
%\keywords{CRC codes \and distance \and linear recurring sequences \and
%  polynomial's ring \and weight distribution}
\end{abstract}

\section{Introduction}

Cyclic Redundancy Check (CRC) codes  are an important class of error
detecting codes. These codes are widely used in computer
communication networks because of their easy and fast encoder and
decoder implementation and their considerable burst-error detection
capability. This properties 
are provided by the structure of shortened cyclic code. This
capability to detect burst-errors is well-studied in \cite{wicker}. 

To measure the degree of goodness of error-detecting codes, we have to
investigate two properties. The first  is
the \emph{minimum distance} of the code. This quantity is the smallest
number of bit positions in which any two given words of the code differ. The
second is the \emph{undetected error probability}
($P_{ue}$) that 
measures the probability that an error occurs during 
transmission that cannot be detected by the decoder. The performance of
the code improves when the minimum distance increases or when $
P_{ue} $ decreases. 

To investigate these two properties, it is important to know the
\emph{weight distribution} of the code. A way to compute this
distribution is to list all of the words of the dual code and compute their
Hamming weights. The \emph{weight distribution} of the code is then
provided  by the Theorem of MacWilliams \cite{mac}. 

The structure of CRC codes offers the opportunity to construct an
\emph{ad-hoc} algorithm that has less computational cost; see
\cite{casta} for a treatment of the binary case.

This work extends the
algorithm to CRC codes over any finite field. 

The second section of this paper is concerned with preliminary
notions. We treat more precisely, but not in detail, CRC
codes and properties they have in common with cyclic codes. A good
working definition of CRC codes is also given. 

The third section deals with the fundamental step of the algorithm. We
examine the connection between Linear Recurring Sequences
(LRS's) and words of the dual code of a CRC code, explain the need of
the choice of the best LRS and  
consider the bijective  relation between LRS's and elements of
$\F_q[x]/(g(x))$, where $g(x)$ is the polynomial
generating the code.

We will then turn our interest to the structure of the ring $
\F_q[x]/(g(x)) $.  The fourth section shows that it is possible to use
the \emph{Chinese Remainder Theorem} in order to work with quotients 
rings via powers of irreducible polynomials.

The main task of the fifth section is to find representatives of the
$x$-orbits in the ring $\F_q[x]/(g(x)^t)$, where $g(x)$ is an
irreducible polynomial. First of all we obtain a decomposition of this
ring in an union of sets which are stable under
$x$-multiplication. First we go deep into the
representation of the multiplicative group of a ring as a product of
cyclic groups. A set of generators of these cyclic
groups is explicitly shown. Then, we use the preceding results 
to construct a set of
representatives of every possible $ x $-orbit of the ring.

\section{Preliminaries}\label{seconda}

In the introduction, we stated that CRC codes are extensively used
nowadays. Despite that in literature there is differing definitions
of this code. This is raised to the different ideas about their
utilize. Now we will give a definition of this codes from
\cite{joachim} to then bring back us to the more operational one.

\begin{defi}
Let $g(x)\in \F_q[x]$ be a monic polynomial over the finite field
$\F_q$ of characteristic $p$. Let us consider the encoding map
\begin{eqnarray*}
\phi:F_q[x] & \rightarrow & F_q[x]\\
m(x) & \mapsto & c(x)=m(x)g(x).
\end{eqnarray*}
A Cyclic Redundancy Check (CRC) code  is then the ideal
$\big(g(x)\big)={\rm im}\; \phi$. 
\end{defi}

This definition give the basic property of CRC codes, i.e. the fact
that they are generated by a \emph{generator polynomial} $g(x)$. 

Such a definition of CRC codes is appropriate from a theoretical point of
view, but in application this definition is not enough. The resulting
code is not \emph{observable}, see \cite{joachim}. 

One way to correct this problem in to predetermine the
length of the message.
This allows the receiver to test for code membership by long division.
If $c(x)$ is the received word, compute
$$c(x)=\tilde{m}(x)g(x)+r(x).$$ 
If $r(x)=0$, then the receiver can conclude that $\tilde{m}(x)$ is the
transmitted message $m(x).$ Otherwise a retransmission will be requested.

We thus arrive at a better working definition of CRC codes.

\begin{defi}
Let $n,r\in \N$ with $n>r>0$. Let $q\in \N$ be some power of a prime
number $p$ and $g(x)\in \F_q[x]$ a monic polynomial such that
$\deg{g(x)}=r$ and $g(0)\not= 0$.

A $(n,n-r)$ CRC code $C$ is the set
$$C=\left\{ c(x)\in \F_q[x] \;|\; c(x)=g(x)m(x),\;
  \deg{m(x)}<n-r\right\}.$$      
\end{defi}

Such a set has the structure of a linear code. 
We note that a CRC code
is a  cyclic code if and only if the generator polynomial $g(x)$
divides $x^n-1$. 

From this representation, it is easy to deduce that CRC codes
are shortened cyclic codes. In fact, a basis of  a CRC code
can be formed by $x$-multiplications of the generator polynomial. 

As previously stated, the Theorem of MacWilliams give us the
possibility to switch our interest to the weight distribution of the
dual code.

The dual code of a CRC code has an interesting structure. Given a
polynomial $ g(x) $ over $ \F_q $, the dual code of a CRC code of any
length generated by $ g(x) $ is isomorphic to the
ring $\F_q[x]/(g(x)).$ 

Via some easy steps which come from the theory of dual codes, it is easy
to deduce the following property of dual codewords:

\begin{prop}\label{dualword}
Let $ C \in \F_q^{n}$ be a CRC code of and
$$g(x)=g_0+g_1x+\dots+g_{r-1}x^{r-1}+x^r$$ its generator
polynomial. Then $c=(c_0,\dots,c_{n-1})$ is an element of
the dual code $C^\perp$ if and only if its components satisfy the
relation 
$$c_i=-g_0c_{i-r}-\dots-g_{r-1}c_{i-1}, \mbox{\ \ \ } i=r,\dots,n.$$
\end{prop}

\subsection*{Notations}

In this work, we will use the following notations:
\begin{itemize}
\item $p$ will be the prime number that is the characteristic of the ring
  $\F_q$; then $q$ is a power of $p$, i.e. $q=p^\delta$ for some
  $\delta \in\N_+$; 
\item $n\in \N_+$ will be the length of the CRC code;
\item $g(x)\in \F_q[x]$ will be the monic generator polynomial of a CRC code,
  with $g(0)\not=0$, $\deg{g(x)}=r$ and $0<r<n$;
\item $g(x)=\prod_{i=l}^m g_l(x)^{e_l}$ will be the irreducible
  decomposition of $g(x)$;
\item $u$ will be an element of the ring $\F_q[x]/(g(x))$ and $u(x)$ the
  representative of lowest degree of $u$ in $\F_q[x]$;
\item $\R_g^q$ will be the ring $\F_q[x]/(g(x))$ and $\R_{g^t}^q$ the ring
  $\F_q[x]/(g(x)^t)$;
\item $M_g^q$ will be the multiplicative group of $\R_g^q$,
  i.e. $\left(\F_q[x]/(g(x))\right)^*$, and  $M_{g^t}^q$ the
  multiplicative group of the ring $\R_{g^t}^q$. 
\end{itemize}

\section{Quotient Ring by a Primitive Polynomial \\
  and Fundamental Step of the Algorithm} 

In this section the fundamental step of the algorithm will be
illustrated. We will use algebraic objects such as linear recurring
sequences (LRS's) \cite{lidl} and polynomials over finite fields.

The next theorem recalls the part of Kronecker's Theorem
\cite{kronecker} which is the most interesting for our purpose. 

\begin{teo}\label{kron}
Let $u(x)\in \F_q[x]$ be a polynomial with $\deg{u(x)}<\deg{g(x)}$. \\
Then there exists exactly one sequence $(c_i)_{i\in \N}\subset
\F_q^\N$ such that
$$\frac{u(x)}{g(x)}=\sum_{i=0}^\infty \frac{c_i}{x^{i+1}}=:c(1/x).$$
Moreover the sequence $(c_i)_{i\in \N}$ satisfies the linear relation
\begin{equation}\label{lrelation}
c_i=-g_0c_{i-r}-\dots-g_{r-1}c_{i-1}, \mbox{\ \ \  } i\geq r.
\end{equation}
\end{teo}

As an immediate application one obtains the following corollary.

\begin{cor}\label{bijection}
There exists a bijection between the ring $\R_g^q$ and the set
of all LRS with characteristic polynomial $g(x)$.
\end{cor}

From the preliminaries stated in the previous section and from Corollary
\ref{bijection}, it follows that there is a bijection between the set of
LRS's with 
characteristic polynomial $g(x)$ and the dual code of any CRC code
whose generator polynomial is $g(x)$. In the following theorem, we
make this bijection explicit.

\begin{teo}
Let $L_g$ be the set of LRS's over $\F_q$ with characteristic
polynomial $g(x)$. Let $C$ be an $(n,n-r)$ CRC code over $\F_q$ whose
generator polynomial is $g(x)$, and $C^\perp$ its dual code. Then the
following relation
\begin{eqnarray*}
\psi: \ \ \ \ \ \ L_g & \rightarrow & C^\perp\\
    (c_i)_{i\in \N} & \mapsto & (c_0,\dots,c_{n-1})
\end{eqnarray*}
is bijective.
\end{teo}

This relation allows one to work with LRS's 
instead of with words of the dual code. Now we want to represent any
word of the dual code through some LRS. We want to use the
minimum possible number of LRS's to representing the dual code. The next
lemma will give us the chance to take only some of the LRS's of $L_g$ in
representing the code $C^\perp$:

\begin{lem}\label{extraction}
Let $C\subset \F_q^n$ be a CRC code with generator polynomial $g(x)$, and
$(c_i)_{i\in \N}\subset \F_q^\N$ a LRS whose characteristic polynomial is
$g(x)$. Then
$$(c_k,\dots,c_{k+n-1})\in C^\perp \ \ \forall\; k\in \N$$
\end{lem}

This Lemma gives us a way to ``extract'' words of the dual code using
only a LRS and the length of the code.

We will now obtain the best way to construct a LRS. We will use the
lowest degree representative element of a class of $\R_g^q$ and divide 
it by the monic polynomial $g(x)$. 

Let now $u(x)\in F_q[x]$ be a polynomial satisfying the
hypothesis of Theorem \ref{kron}. A method for obtaining the
related LRS is explained in \cite{casta}. The method
follows from the relation
\begin{equation}\label{relation}
\frac{u(x)}{g(x)}=\frac{u_{r-1}}{x}+\frac{u'(x)}{xg(x)},
\end{equation}
where $u_{r-1}$ is the coefficient of the $(p-1)$-th degree term of  the
polynomial $u(x)$ and $u'(x)=xu(x)-u_{r-1}g(x)\equiv xu(x)\pmod{g(x)}.$ 
It is trivial to see that the polynomial $u'(x)$ satisfies the
hypothesis of  Theorem \ref{kron}
as well. Relation \eqref{relation} can be iterated,
and the resulting sequence of coefficients $u_{r-1}$ is a LRS. 

An interesting remark is that LRS can be easily constructed by
using a Linear Feedback Shift Register (LFSR) with generator polynomial
$g(x)$.
The next subsection is devoted to the fundamental step of the algorithm.

\subsection{The Fundamental Step}

Let us now consider a LRS $(c_i)_{i\in \N}$ with characteristic
polynomial $g(x)$. We will make the way to extract words of the
dual code of a $(n,n-r)$ CRC code explicit. The following figure
depicts the idea of algorithm; this scheme
follows from Lemma \ref{extraction}.
   
\begin{figure}[h]
  \psfrag{c_0}{\hspace{1pt}\scalebox{.6}{$c_0$}}
  \psfrag{c_1}{\hspace{1pt}\scalebox{.6}{$c_1$}}
  \psfrag{c_2}{\hspace{1pt}\scalebox{.6}{$c_2$}}
  \psfrag{c_3}{\hspace{1pt}\scalebox{.6}{$c_3$}}
  \psfrag{c_n-1}{\hspace{-1.5pt} \scalebox{.6}{$c_{n-1}$}}
  \psfrag{c_n}{\hspace{1.5pt}\scalebox{.6}{$c_{n}$}}
  \psfrag{c_n+1}{\hspace{-2.5pt} \scalebox{.6}{$c_{n+1}$}}
  \psfrag{c_n+2}{\hspace{-1pt} \scalebox{.6}{$c_{n+2}$}}
  \psfrag{c_n+3}{\hspace{-1pt} \scalebox{.6}{$c_{n+3}$}}
  \psfrag{c^0}{\scalebox{.6}{$c^{(0)}$}}
  \psfrag{c^1}{\scalebox{.6}{$c^{(1)}$}}
  \psfrag{c^2}{\scalebox{.6}{$c^{(2)}$}}
  \includegraphics[scale=.60]{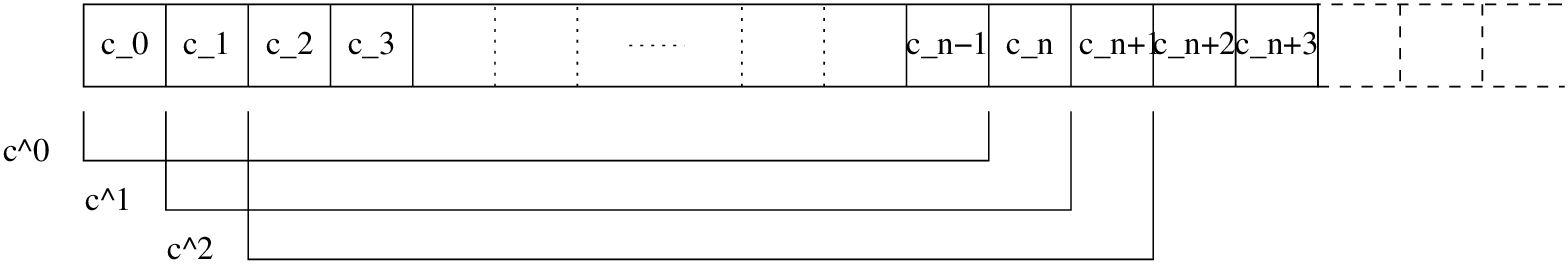}
\end{figure}

In the figure above, $c^{(k)}\subset \F_q^n$ denotes the $k$-th word
of the dual code extracted from the above sequence.

The figure leads directly to relations between the weight distribution
of the words thus extracted:

\begin{oss}(Weight relations between words) 
\begin{itemize}
\item if $c_{k-1}\not=0$ and $c_{k+n-1}=0$, then
  $wt(c^{(k)})=wt(c^{(k-1)})-1$;
\item if $c_{k-1}=0$ and $c_{k+n-1}\not=0$, then
  $wt(c^{(k)})=wt(c^{(k-1)})+1$;
\item $wt(c^{(k)})=wt(c^{(k-1)})$ otherwise.
\end{itemize}
\end{oss}  

This remark will be very useful in decreasing the
computational cost of the algorithm. In such a way, once the
weight of the first word extracted from a LRS $(c_i)_{i\in \N}$ has
been computed, the weights of the following words  can be easily
determined. This procedure has a minimal computational cost, cause the
operations of addition or subtraction are constant time complexity
operations. 

In \cite{casta}, a way to compute these weights from the 
same two LFSR is also explained. The second LFSR has
to be shifted $ n $ times. The first LFSR will give
the new input component (bit) of the constructed word and the second one the
leaving one. 

\subsection{Relation Between LRS's and Words of $C^\perp$}

Now we are able to extract words of the dual code from LRS's, but
some questions are still
unanswered. What is a minimal set of LRS's sufficient to determine the
weight distribution? How can we be sure that we are not
considering the same word more than once? 

We define next sets:

\begin{defi}
Let $u(x)\in \F_q[x]$ with $\deg{u(x)}<\deg{g(x)}$, and $(c_i)_{i\in
  \N}\subset \F_q^\N$ be the LRS
related to $u(x)$ (see Theorem {\rm \ref{kron}}). We denote with
$C^\perp_{u}\subset C^\perp$ the set of all words of the dual code of a
CRC code extracted from  $(c_i)_{i\in \N}$.
\end{defi}

The next Lemma states a bound on the number
of different words that can be extracted from a fixed LRS. 

\begin{lem}
Let $u(x)\in \F_q[x]$ such that $\deg{u(x)}<\deg{g(x)}$. The cardinality
of $C^\perp_{u}$ is
$$|C^\perp_{u}|=\mbox{\emph{ord}}\left(
  \frac{g(x)}{\gcd(g(x),u(x))}\right).$$    
\end{lem} 

The proof of Lemma follows directly from the relation between the
number of words that can be extracted and the
period of $ (c_i)_{i\in \N} $ and the definition of the \emph{order} of a
polynomial. Details can be found in \cite{casta}.   

\begin{defi}
Let $u(x)\in \F_q[x]$, the order of $u(x)$ is the least natural number
$o_u$ such that $u(x)$ divides $x^{o_u}-1$.
\end{defi}

\subsection{$x$-orbits, LRS and Words of $C^\perp$}

In order to continue we need another algebraic structure, i.e. the
\emph{$x$-orbits} of the ring 
$\R_g^q$. As we have already remarked, the $g(x)$ and $x$ are
relatively prime. Let $(\mbox{x})$ denote the cyclic
subgroup of the 
multiplicative group $M_g^q$ generated by $ x $, i.e.
$$(\mbox{x}):=\left\{x^k\in M_g^q \;|\; k\in \N\right\}.$$
  
\begin{defi}
The $x$-orbits are the sets resulting from the action of the cyclic
group $(\mbox{\emph{x}})$ on the ring $\R_g^q$. For $ u $ an element
of $\R_g^q$,  we denote with $\mathfrak{C}^\perp_{u}$ the
$x$-orbit of $u$. 
\end{defi}

It is well known that $x$-orbits can be considered as equivalence classes
of the ring $\R_g^q$. 

The next lemma gives an explicit relation between the $x$-orbits of
two distinct elements of
$\R_g^q$ and the respectively generated LRS's. The lemma will work with sets
of words $C^\perp_{u}$, i.e. words extracted from a LRS.

\begin{lem}
Let $u_1,u_2$ be two distinct elements of $\R_g^q$. Then the following relation
$$u_2\in \mathfrak{C}^\perp_{u_1} \iff C^\perp_{u_1}=C^\perp_{u_2}$$ 
holds.
\end{lem}
\begin{dem}
The proof consists of two parts.
\begin{itemize}
\item[$(\Rightarrow)$] By definition $u_2$ belongs to
  $\mathfrak{C}^\perp_{u_1}$ if and only if there exists $j\in \N$
  such that 
  $$u_2(x)\equiv x^ju_1(x)\pmod{g(x)}.$$
  Hence the LRS constructed from the polynomial $u_2(x)$ is the same as
  the LRS obtained by shifting that of the first
  polynomial $ j $ times, so that $C^\perp_{u_2}\subset C^\perp_{u_1}$.\\
  The cardinality of the two sets is the same, as
  \begin{eqnarray*}
    |C^\perp_{u_2}|&=&\mbox{ord}\left(
      \frac{g(x)}{\gcd(g(x),u_2(x))}\right)=\mbox{ord}\left(
      \frac{g(x)}{\gcd(g(x),x^ju_1(x))}\right)=\\
    &=&\mbox{ord}\left(
      \frac{g(x)}{\gcd(g(x),u_1(x))}\right)=|C^\perp_{u_1}| 
  \end{eqnarray*}
  since $\gcd(x,g(x))=1$. This implies $C^\perp_{u_2}=C^\perp_{u_1}$.
\item[($\Leftarrow$)] Let us suppose that there is no $j\in \N$
  such that $$u_2(x)\equiv x^ju_1(x)\pmod{g(x)}.$$ This implies that
  in the development of 
  relation \eqref{relation} beginning with $u_1(x)$, the polynomial
  $u_2(x)$ cannot be found in the right-hand side of the relation
  \eqref{relation}. Hence the LRS related to the second polynomial 
  cannot be obtained as a shift of the LRS related to the first
  one. This implies that the sets $C^\perp_{u_1}$ and
  $C^\perp_{u_2}$ are different.
\end{itemize}
The proof is complete.
\end{dem}

Previously, in Section \ref{seconda}, we stated the bijective relation between the dual code of a
CRC code and the ring $\R_g^q$. It follows from the previous lemma that
the dual code can be constructed by taking the union of disjoint sets that are
related to the $x$-orbits of the ring $\R_g^q$. These orbits are
also related to LRS's. Our goal is to find a representative of each
$x$-orbit. Thereafter using the
\emph{fundamental step} we will be able to compute the
weight distribution of the dual code.

\section{Application of the Chinese Remainder Theorem}

We want now to 
obtain a representation of the structure of the ring $\R_g^q$
that will be useful for our particular algorithm. We will look for
a decomposition of the ring into $ x $-orbits.

From the Chinese Remainder Theorem we know that
\begin{equation}\label{irred_decomp}
\R_g^q\approx \prod_{l=1}^m\R_{g_l^{e_l}}^q,
\end{equation}
where $g(x)=\prod_{l=1}^mg_l(x)^{e_l}$ is the irreducible
factor decomposition. 

Let us write the isomorphism explicitly in our case. The following theorem is
claimed implicitly in \cite{casta}.

\begin{teo}
Let $g(x)\in \F_q[x]$ be a monic polynomial such that $g(0)\not=0$, and
let us consider its irreducible decomposition.

The map
$$\phi:\R_g^q \rightarrow
\prod_{l=1}^m\R_{g_l^{e_l}}^q$$ 
given by $\phi(u)=(u_1,\dots,u_m)$ with
$$u_l(x)\equiv u(x)\pmod{g_l(x)^{e_l}}$$
is an isomorphism with inverse 
$$\phi^{-1}(u_1,\dots,u_m)=\sum_{l=1}^mu_l(x)v_l(x)
  \frac{g(x)}{g_l(x)^{e_l}} \pmod{g(x)},$$
where $v_l(x)$ is the multiplicative inverse of $g(x)/g_l(x)^{e_l}$ in
$\R_{g_l^{e_l}}^q$. 
\end{teo}

Thanks to this theorem  we can
begin our  study in the case of quotient rings of powers of
irreducible polynomials. 

Let $u_l$ be an element of the ring $\R_{g_l^{e_l}}^q$; the result of
the action of $ x^{k_l} $ on $ u_l \in \R_{g_l^{e_l}}^q $ will be
denoted by $ u_l^{(k_l)} $,  i.e. $u_l^{(k_l)}\equiv x^{k_l}u_l
\pmod{g_l(x)^{e_l}}$. 
From the paper \cite{casta} and some calculations, one obtains
the next theorem.

\begin{teo}\label{TCRappl}
Let $g(x)\in \F_q[x]$ be monic with
  $g(0)\not=0$ and $g(x)=\prod_{l=1}^mg_l(x)^{e_l}$ be its
irreducible decomposition.  Let also $\mathfrak{C}^\perp_{u_l}$ be
the $x$-orbits of $\R_{g_l^{e_l}}^q$ for \\$l=1,\dots,m$, with
cardinality $d_l$ and representative $u_l$ respectively.

It follows that representatives of any $x$-orbits of the ring $\R_g^q$ are
$$\left(u_1,u_2^{(k_2)},\dots,u_m^{(k_m)}\right)\in \prod_{l=1}^m
\R_{g_l^{e_l}}^q$$ 
for $0\leq k_l <\mathcal{K}_l$, for  $l=2,\dots,m$ and
$$\mathcal{K}_l=\gcd (d_l, \mbox{\emph{lcm}}(d_1,\dots,d_{l-1})).$$
\end{teo}

A proof of the previous theorem can be done by induction using the
paragraph titled \emph{Action of a Cyclic Group on a Cartesian
  Product} of \cite{casta}, where the authors analyze the case of the
cartesian product of two sets.

\section{Decomposition of $\R_{g^t}^q$ into $x$-orbits}

Let $t$ be a natural number and let us consider an irreducible  polynomial
$g(x)\in \F_q[x]$ of degree $r$. The following representation of the
elements of $\R_{g^t}^q$ is also valid and will be useful for our work. 

\begin{lem}\label{gadically}
Every $f\in \R_{g^t}^q$ can be represented in an unique way as
$$f=\left[\sum_{l=0}^{t-1} f_l(x)g(x)^l\right],$$
where $f_l(x)\in \F_q[x]$ and $\deg{f_l(x)}<\deg{g(x)}$. 
\end{lem}

This representation of an element of $\R_{g^t}^q$ follows directly
from the representation of any representative of the class in $g(x)$
base. 

The next step is to investigate subsets of the ring
$\R_{g^t}^q$ which are closed under $x$-multiplication. The following
theorem gives information about such subsets.

\begin{teo}
Let $u(x)\in \F_q[x]$ be the representative of minimal degree of a class
$u\in \R_{g^t}^q$, and let $s$ be the natural number  
$$s:= \max \left\{i\in \N \;\big|\; g(x)^i|u(x)\right\}.$$
If we denote by $\bar{u}(x)\in \F_q[x]$ the polynomial obtained by
dividing $u(x)$ by $g(x)^s$, then the following relation holds:
\begin{equation*}
u'\in \mathfrak{C}^\perp_{u} \subset \R_{g^t}^q \iff \left\{
\begin{array}{l}
\max \left\{i\in \N \;\big|\; g(x)^i|u'(x)\right\} = s, \\ \\
\left[u'(x)/g(x)^s\right] \in \mathfrak{C}^\perp_{\bar{u}}\subset
\R_{g^{t-s}}^q. 
\end{array}
\right.
\end{equation*}
\end{teo}

The proof of this theorem is an easy computation.

\begin{oss}
From the previous theorem, two remarks can be extracted: 
\begin{enumerate}
\item give an $x$-orbit $\mathfrak{C}^\perp_{u}$, the maximal power of
  $g(x)$ that divides an element of the orbit does not depend of the
  choice of the element; 
\item the choice of the best representative of $x$-orbits can be limited
  to the set  of elements of $\R_{g^t}^q$ whose representatives in $
  \F_q[x] $ are coprime with $g(x)$.
\end{enumerate}
\end{oss}

The next corollary follows from Lemma \ref{gadically} and the previous
theorem.   

\begin{cor}\label{decomposition}
The ring $\R_{g^t}^q$ can be decomposed as follows:
$$\R_{g^t}^q=\{0\}\cup \bigsqcup_{l=1}^{t-1} g(x)^l\cdot
M_{g^{t-l}}^q,$$
where the sets 
$$g(x)^l\cdot M_{g^{t-l}}^q=\left\{u\in\R_{g^t}^q
  \;\big|\; u=[g(x)^l\bar{u}(x)],\; \bar{u}\in M_{g^{t-l}}
\right\}$$
are stable under $x$-multiplication. 
\end{cor}

\subsection{Characterization of elements of  $M_{g^l}^q$}

Let us initially give a corollary of Lemma \ref{gadically}.

\begin{cor}\label{invertible}
The element $f\in \R_{g^t}^q$ is invertible if and only if, in the representation
given in Lemma \ref{gadically}, $f_0(x) \not=0$.
\end{cor}

The group $M_{g^l}^q$ is the multiplicative group of the ring
$\R_{g^l}^q$ and is finite. By the theory of finitely generated
abelian groups, $M_{g^l}^q$ can be expressed as a
product of cyclic groups. Let us investigate this
structure more precisely. 

Let us distinguish between two cases: $l=1$ and $l\geq 2$.
The first case should also be split into two parts: either $g(x)$ is
primitive, or it is not. 
In both cases, the ring $\R_q^q$ is also a finite field, hence its
multiplicative group is cyclic. The difference between the primitive
and the non-primitive case lies in the choice of the generator
element. If $g(x)$ is a primitive polynomial, i.e. if
$\ord{g(x)}=q^r-1$, where $\deg{g(x)}=r$, then $x\in M_{g}^q$ is a
good choice of generator. 
Otherwise, $x$ is not a generator anymore. We will denote
by $h\in M_g^q$ a generator of the group. 

Let us now consider the case $l\geq 2$. 

\begin{teo}
The order of the group $M_{g^l}^q$ is $(q^r-1)q^{(l-1)r}$. Moreover 
$$M_{g^l}^q\approx M_g^q\times S_p,$$
where $S_p$ is the $p$-Sylow subgroup of $M_{g^l}^q$.
\end{teo}

The group $M_g^q$ has already been analyzed; we now study the structure
of the $p$-Sylow subgroup.

\begin{teo}\label{p-elements}
Let $f\in M_{g^l}^q$. The multiplicative order of $f$ is a power of
$p$ if and only if there exists a polynomial $m(x)\in \F_q[x]$ such
that 
$$f=[1+m(x)g(x)].$$
\end{teo}

\begin{dem}
Let $f(x)\in \F_q[x]$ be the representative of lowest degree of
$f$, where $f$ is an element whose order is a power of $p$. There
exists a unique way to write 
$$f(x)=f_0(x)+f^{(1)}(x)g(x)$$ 
with  $\deg{f_0(x)}<\deg{g(x)}$.

Let $k\in \N_+$ be such that $p^k>l$. Then
$$f(x)^{p^k}=(f_0(x)+f^{(1)}(x)g(x))^{p^k}=
f_0(x)^{p^k}+(f^{(1)}(x)g(x))^{p^k}.$$
In $\R_{g^l}^q$, this relation reduces to $f^{p^k}=[f_0(x)^{p^k}]$, and
$f_0(x)^{p^k}$ is the lowest-degree representative of the
class. From the equality criteria between polynomials, we see that
$f_0(x)=1$. 

Vice-versa, let $m(x)\in \F_q[x]$ be a polynomial such that 
$$f(x)=1+m(x)g(x).$$
Let $k\in \N_+$ be such that $p^k>l$; then
$$f(x)^{p^k}=(1+m(x)g(x))^{p^k}=
1+(m(x)g(x))^{p^k},$$
and this represents the identity in $ \R_{g^l}^q $.
\end{dem}

\begin{oss}
From now on we will denote by $\A$ an element algebraic over $\F_p$ of
degree $\delta$. Recall that $ \delta $ is such that $q=p^\delta$. Therefore, we obtain $\F_q\approx \F_p[\A]$. 
\end{oss}

Another notation that we will use extensively is the following:
\begin{equation}\label{condizioni}
a_{i,j,k}(x)= 1+\A^ix^jg(x)^k\in S_p,
\end{equation}
where $0\leq i < \delta$, $0\leq j < r$ and $1\leq
k<l$. We are now able to state the theorem that specifies the decomposition
of the $p$-Sylow subgroup into a product of cyclic groups.

\begin{teo}\label{generatori}
Let $S_p$ be the $p$-Sylow subgroup of $M_{g^l}^q$. The
following isomorphism holds:
$$S_p\approx \prod_{i,j,k} \big(a_{i,j,k}(x)\big),$$
where $\big(a_{i,j,k}(x)\big)\subset S_p$ is the cyclic group generated by
$a_{i,j,k}(x)$ and the parameter $k$ satisfies the 
condition $p\nmid k$. 
\end{teo}

The proof of this theorem follows from the next lemma.

\begin{lem}
For any polynomial $$f(x)=1+f_h(x)g(x)^h+m(x)g(x)^{h+1}\in \F_q[x]$$
with $h\in \N_+$ and $\deg{f_h(x)}<\deg{g(x)}$, there exist numbers
$$c_{(i,j)}\in \{0,1,\dots,p-1\}$$ such that
$$\prod_{i,j}a_{i,j,h}(x)^{c_{(i,j)}}\equiv f(x) \pmod{g(x)^{h+1}}$$
for every $0\leq i < \delta$ and $0\leq j < r$.  
\end{lem}

\begin{dem}
The proof will be split into two parts: either $ p \nmid h $, or $ p
\; | \; h $. 

Let us first suppose that $p\nmid h$. The polynomial $f_h(x)$ can be
written as
$$f_h(x)= \sum_{i,j}c_{(ij)}\A^ix^j,$$
with $c_{(ij)}\in \F_p.$ It follows that
\begin{eqnarray*}
\prod_{i,j}a_{i,j,h}(x)^{c_{(ij)}} & = &
1+\sum_{i,j}c_{(ij)}\A^ix^jg(x)^h+ m'(x)g(x)^{h+1}\equiv\\
&\equiv& f(x) \pmod{g(x)^{h+1}}.
\end{eqnarray*}

Otherwise, if $p\;|\; h$ then $h=ph'$. The polynomial $g(x)$ is
irreducible, thus the field $\R_g^q$ is perfect. This implies that the
projection 
$f_h\in \R_g^q$ of $f_h(x)$ is the $p$-th power of
some element $ l \in \R_g^q$.
Then we have
$$l(x)^p\equiv f_h(x)\pmod{g(x)},$$
where $l(x)$ is the representative of $l$ of lowest degree and then 
$$l(x)=\sum_{i,j}c_{(ij)}\A^ix^j\in \F_q[x].$$
Hence, we can conclude that
\begin{eqnarray*}
\big(\prod_{i,j}a_{i,j,h'}(x)^{c_{(ij)}}\big)^p&=&\big(1+l(x)g(x)^{h'}+
\tilde{l}(x)g(x)^{h'+1}\big)^p=\\
&=& 1+l(x)^pg(x)^{ph'}+\bar{l}(x)g(x)^{p(h'+1)}\equiv f(x)\pmod{g(x)^{h+1}},
\end{eqnarray*}
and the proof is complete.
\end{dem}

\begin{dem}[Theorem \ref{generatori}]
The polynomial $f_1(x)$ can be expressed as
$$f_1(x)=\sum_{i,j}c_{(ij,1)}\A^ix^j.$$
From the previous lemma we have
\begin{equation}\label{parte1}
\prod_{i,j}a_{i,j,1}(x)^{c_{(ij,1)}}=1+f_1(x)g(x)+m(x)g(x)^2.
\end{equation}
Let now consider $\tilde{f}_2(x)\equiv f_2(x)-m(x)\pmod{g(x)}$; then
$$\tilde{f}_2(x)=\sum_{i,j}c_{(ij,2)}\A^ix^j.$$
Using the lemma once more, we obtain
\begin{equation}\label{parte2}
\prod_{i,j}a_{i,j,2}(x)^{c_{(ij,2)}}=1+\tilde{f}_2(x)g(x)^2+
\tilde{m}(x)g(x)^4.
\end{equation} 
Let us now multiply the relations \eqref{parte1} and
\eqref{parte2}:
\begin{eqnarray*}
&& (1+f_1(x)g(x)+m(x)g(x)^2)(1+\tilde{f}_2(x)g(x)^2+
\tilde{m}(x)g(x)^4)=\\
&& = 1+f_1(x)g(x)+f_2(x)+\hat{m}(x)g(x)^3.
\end{eqnarray*}

The claim is obtained by iterating this computation $ l $ times.
\end{dem}

Let us now consider the homomorphism of groups 
$$\mu:\prod_{i,j,k}\left(a_{i,j,k}(x)\right)\rightarrow S_p,$$
where the parameters satisfy the conditions given in \eqref{condizioni} and
in addition $p\nmid k$.
To prove that the map above is an isomorphism it is enough to prove
injectivity. 

\begin{teo}
With the conditions previously given on the parameters $i,j,k$, the following
holds:
$$\prod_{i,j,k}a_{i,j,k}(x)^{c_{(ijk)}}\equiv 1 \pmod{g(x)^l}\iff
c_{(ijk)}\equiv 0 \pmod{\ord{a_{i,j,k}(x)}}.$$
\end{teo} 

\begin{dem}
Let us begin by expanding the power of each polynomial $a_{i,j,k}(x)$.
Writing the exponents as
$$c_{i,j,k}=p^{s_{(ijk)}}c'_{(ijk)},$$
it follows that
\begin{eqnarray}
a_{i,j,k}(x)^{c_{(ijk)}}&=&\left( 1+\A^ix^jg(x)^k \right)^{c_{(ijk)}}\nonumber\\
&=& \left(
  1+\A^{ip^{s_{(ijk)}}}x^{jp^{s_{(ijk)}}}g(x)^{kp^{s_{(ijk)}}}\right)
^{c'{(ijk)}}\nonumber\\
&=& 1+\sum_{h=1}^{c'_{(ijk)}} \binom{c'_{(ijk)}}{h} \left(
  \A^{ip^{s_{(ijk)}}}x^{jp^{s_{(ijk)}}}g(x)^{kp^{s_{(ijk)}}}\right)^h.  
\label{spezzettato}    
\end{eqnarray}

The next step is to find the minimum exponent of $g(x)$ in
\eqref{spezzettato}. We have to highlight all the terms where this exponent
occurs to continue with the proof. Let us introduce the notation 
$$\mathfrak{K}p^{\mathfrak{s}}:=\min_{i,j,k}kp^{s_{(ijk)}}.$$
In order to use this notation it is important that $p\nmid k$, so
that any triplet $(\bar{i},\bar{j},\bar{k})$ giving this minimum is
such that $s_{(\bar{i}\bar{j}\bar{k})}=\mathfrak{s}$.

Now we will deal with the product
$\prod_{i,j,k}a_{i,j,k}(x)^{c_{(ijk)}}$. If we use 
\eqref{spezzettato} and group all monomials according to the power
of $ g(x) $  they contain, we obtain  
$$m_{\mathfrak{K}p^{\mathfrak{s}}}(x):=\sum_{(i,j)\in \mathcal{J}}
c'_{(ijk)}\A^{ip^{\mathfrak{s}}}x^{jp^{\mathfrak{s}}}=\big(\sum_{(i,j)\in \mathcal{J}}
c'_{(ijk)}\A^{i}x^{j}\big)^{p^{\mathfrak{s}}}$$ 
where $\mathcal{J}$ is the set of all pairs $(i,j)$ for which the
exponent $kp^{s_{(ijk)}}$ is minimal.    

The polynomial $m'(x):=\sum_{(i,j)\in \mathcal{J}} c'_{(ijk)}\A^{i}x^{j}$
does not vanish, because the pairs $(i,j)$ appear once in the sum.
Moreover, the condition $j<\deg{g(x)}$ says that no factor of $g(x)$
divides $m'(x)$. The hypothesis is then satisfied if and only if
$\mathfrak{K}p^{\mathfrak{s}} \geq l$.

This last remark concludes the proof, because for any single factor of
the product it follows that 
$$a_{i,j,k}(x)^{c_{(ijk)}}\equiv 1 \pmod{g(x)^l},$$
so that $c_{(ijk)}\equiv 0 \pmod{\ord{a_{i,j,k}(x)}}.$
\end{dem}

\subsection{Set of generators of the $x$-orbits of $M_{g^l}^q$}

In this subsection we will make the representatives of the $x$-orbits
of the ring explicit. We will use the results of previous sections to express
these representatives via generators of the groups $M_{g^l}^q$ with $l<t$.
In Corollary \ref{decomposition} we saw how to decompose the ring into
a disjoint union of sets stable under $x$-multiplication. The
representatives are then to be looked for in these sets.

Let us give the order of the generators of the $p$-Sylow subgroup of
$M_{g^l}^q$.

\begin{teo}
Let $l\in \N$. The elements $a_{i,j,k}(x)\in M_{g^l}^q$, with
parameters satisfying \eqref{condizioni} and such that $p\nmid k$,
have order
$$\ord{a_{i,j,k}(x)}=p^{\lceil \log_p{l/k} \rceil}.$$
\end{teo}

\begin{dem}
Thanks to Theorem \ref{p-elements}, the order of the elements
$a_{i,j,k}(x)$ is a power of $p$. For $m\in \N$ we have 
$$a_{i,j,k}(x)^{p^m}=(1+\A^ix^jg(x)^k)^{p^m} \equiv
1+(\A^ix^jg(x)^k))^{p^m} \pmod{g(x)^l}.$$
Since $\gcd (\A^ix^j,g(x))=1$ we get
\begin{eqnarray*}
(\A^ix^jg(x)^k)^{p^m}\equiv 0\pmod{g(x)^l} & \iff & g(x)^{kp^m}\equiv
0 \pmod{g(x)^l}\\
& \iff & kp^m \geq l.
\end{eqnarray*}
It follows that the lowest exponent of $p$ which fulfills the previous
relation is
$$m=\lceil \log_p{l/k} \rceil$$ 
and the proof is complete.
\end{dem}

To investigate  the construction of the $x$-orbits more deeply, we have to
divide the cyclic group $\big(\mbox{x}\big)$ into product of others
cyclic groups.

\begin{teo}
Let $l\in \N$. The order of the element $x\in M_{g^l}^q$ is 
$$\ord{x}=\ord{g(x)}\cdot p^{\lceil \log_p{l}\rceil}.$$
\end{teo}

\begin{dem}
$\ord{x}$ is such that
\begin{equation}\label{ordine}
x^{\ord{x}}\equiv 1 \pmod{g(x)^l} \mbox{ \ \ } \Rightarrow \mbox{ \
  \ } g(x)^l \;|\;  x^{\ord{x}}-1.
\end{equation}

If we write $\ord{x}=m=p^s\bar{m}$, with $p \nmid
\bar{m}$, then
$$x^m-1=x^{p^s\bar{m}}-1=(x^{\bar{m}}-1)^{p^s}.$$
We are working in characteristic $p$, hence every irreducible factor of 
$x^m-1$ has multiplicity $p^s$. Relation \eqref{ordine} tells us that the
multiplicity has to be at least $l$, then $p^s\geq l$, i.e. $s \geq
\lceil \log_p{l}\rceil$.

We also know that 
$$g(x)\;|\; x^m-1 \iff \ord{g(x)}\;|\; m.$$ 
The order of $x$ is the least common multiple of $p^{\lceil
  \log_p{l} \rceil}$ and $\ord{g(x)}$; as these are coprime, and the
claim follows. 
\end{dem}

\begin{cor}
The cyclic group $\big(\mbox{\emph{x}}\big)\subset M_{g^l}^q$ is
isomorphic to 
the product of two cyclic groups whose orders are $p^{\lceil \log_p{l}
  \rceil}$ and $\ord{g(x)}$, respectively.
\end{cor}

\begin{oss}
Using the representation of the groups $M_{g^l}^q$ given in the previous
section we can make the previous corollary explicit. We have
$$\big(x \big) \approx \big(\mbox{x}_p(x)
\big)\times\big(\mbox{x}_{o_g}(x) \big)$$ 
where $\big(\mbox{x}_p(x)\big)\subset S_p$ and
$\big(\mbox{x}_p(x)\big) \subset M_g^q.$
Without loss of generality we can take 
$$\mbox{x}_p(x):= x^{\ord{g(x)}}\in S_p \mbox{ and }
\mbox{x}_{o_g}(x):= h(x)^{\frac{q^r-1}{\ord{g(x)}}}\in M_g^q$$
where $h(x)$ is the generator of $M_q^q$.
\end{oss}

\begin{teo}\label{penultimo}
Let $\mbox{\emph{x}}_p(x)\in S_p \subset M_{g^l}^q$ be an element of
order $p^{\lceil \log_p{l} \rceil}$. There exist parameters $0\leq i_0 <
\delta -1$ and $0\leq j_0 < \deg{g(x)}$ such that 
$$S_p \approx  \big(\mbox{\emph{x}}_p(x) \big)\times
\prod_{\begin{subarray}{c}i,j,k \\(i.j.k)\not=(i_0,j_0,1)\end{subarray}}\big(a_{i,j,k}(x)\big),$$
where $(i,j,k)$ satisfy \eqref{condizioni} and $p\nmid k$.
\end{teo}

\begin{dem}
The element $\mbox{x}_p$ belongs to the $p$-Sylow subgroup; hence, thanks to
Theorem \ref{generatori}, it has a representation via
polynomials $a_{i,j,k}(x)$, i.e.
$$\mbox{x}_p(x)\equiv \prod_{i,j,k}a_{i,j,k}(x)^{c_{(ijk)}}
\pmod{g(x)^l}.$$  
The order of the element is the highest possible power of $ p
$. Thanks to the above representation, there exist $0\leq i_0 < \delta
-1$ and $0\leq j_0 < 
\deg{g(x)}$ such that
$$\gcd (c_{(i_0j_01)},p)=1.$$
For every $e_{(ijk)}\in \N$ the following relation holds
$$\prod_{i,j,k}a_{i,j,k}(x)^{e_{(ijk)}}=\mbox{x}_p(x)^{\tilde{e}_p}\cdot
\prod_{\begin{subarray}{c} i,j,k \\
    (i,j,k)\not=(i_0,j_0,1) \end{subarray}}a_{i,j,k}(x)^{\tilde{e}_{(ijk)}},$$
with 
\begin{eqnarray*}
\tilde{e}_p & \equiv & e_{(i_0j_01)}\cdot (c_{(i_0j_01)})^{-1}
\pmod{p^{\lceil \log_p{l} \rceil}} \\
\tilde{e}_{(ijk)} & \equiv & e_{(ijk)}-c_{(ijk)}\tilde{e}_p
\pmod{p^{\lceil \log_p{l/k} \rceil}}.
\end{eqnarray*}
The proof is complete.
\end{dem}

We can now state the theorem that specifies every possible
representative of the $x$-orbits of $M_{g^l}^q$. 

\begin{teo}\label{ultimo}
Let $g(x)\in \F_q[x]$ be a degree-$ r $ irreducible polynomial and $l\geq 2$.
There exist $0\leq i_0 < \delta -1$ and $0\leq j_0 <
\deg{g(x)}$ such that the set 
$$\Big\{h(x)^t\cdot \prod_{\begin{subarray}{c}(i,j,k)\not=(i_0,j_0,1)\\
  0\leq i < \delta, \ 0\leq j < r\\ 1\leq k \leq t, \ p\nmid k
\end{subarray}}
(1+\A^ix^jg(x)^k)^{c_{(ijk)}}\pmod{g(x)^l}\Big\},$$
is a family of repretatives of orbits in
$M_{g^l}^q$. Here, $h(x)$ is a primitive element of  $\R_{g}^q$, $\A\in
\F_q$ is an algebraic element of degree $\delta$ over $\F_p$,
and $t$ and $c_{(ijk)}$ are such that 
\begin{itemize}
\item $0\leq t< \frac{q^r-1}{\ord{g(x)}}$
\item $0\leq c_{(ijk)}\leq p^{\lceil \log_p{l/k}\rceil}.$
\end{itemize}
\end{teo}

\begin{dem}
The theorem is quickly proved by rewriting
$$u'\in \mathfrak{C}_{u}^\perp \iff \exists\; \nu \in \N \; :
u'(x)\equiv x^\nu u(x)\pmod{g(x)^l}$$
as
$$u'\in \mathfrak{C}_{u}^\perp \iff \exists\; \nu,\mu \in \N \; :
u'(x)\equiv \mbox{x}_p(x)^\nu\mbox{x}_{o_g}(x)^\mu u(x)\pmod{g(x)^l}$$
and applying Theorem \ref{penultimo}.
\end{dem}

\subsection{Backward Steps}
In this part of the article we make some steps backward. 

Our goal was
to construct  the representatives of the ring  $\R_{g}^q$ with $g(x)$
any polynomial of $\F_q[x]$.

Using Theorem \ref{ultimo} we obtain, for every irreducible
polynomial $g_l(x)$ in the decomposition of $g(x)$, the
representatives of their $x$-orbits in $M_{g_l^s}^q$ for $1\leq s<e_l$.
Corollary \ref{decomposition} gives us the opportunity to compute the
representatives of the ring $\R_{g_l^{e_l}}^q$. This step is easily
done: the representatives of $M_{g_l^s}^q$ for $1\leq s<e_l$
multiplied with a well-chosen power of $g(x)$ give as result
representatives of $\R_{g_l^{e_l}}^q$.

By considering all possible irreducible polynomials of the
decomposition of $g(x)$, we obtain representative of all rings
$\R_{g_l^{e_l}}^q$. The next step is to use Theorem \ref{TCRappl},
and with that we make the representative of the ring $\R_g^q$
explicit.  

In involving, at the end, the fundamental step of the algorithm to the
LRS's related to the representatives of $\R_g^q$, we obtain the weight
distribution of the dual code of a CRC code.

\section{Conclusions}

We analyzed the complexity of the algorithm in the case of
$\R_{g^t}^q$ where $g(x)$ is an irreducible polynomial. It turns out
to be 
$$\mathcal{O}(p^{\delta r}\delta rt^2(n+pt)),$$
but since $pt$ is smaller then $n$ in any practical application the
complexity can be reduced to 
$$\mathcal{O}(p^{\delta r}\delta rt^2n).$$
If we consider all the elements of the ring
$\R_{g^t}^q$ instead of only some representatives, the complexity is
$$\mathcal{O}(p^{\delta rt}n).$$
It is easy to see that the two complexities differ only with respect
to the $t$ parameter. The complexity in our case  
is polynomial in $t$, while in the other case it is exponential.

\subsection*{Acknowledgments}

The author is grateful to Patrizia Gianni and Barry
M. Trager for their guidance during this work.

\end{document}